\documentclass{PoS}

\title{Detection of tau neutrinos  by  Imaging Air  Cherenkov Telescopes}

\ShortTitle{Detection of tau neutrinos  by IACTs}

\author{\speaker{Dariusz G\'ora} \\
        Institut f\"ur Physik, Humboldt Universit\"at, Newtonstr. 15,  D-12489 Berlin, Germany\\
        E-mail: \email{Dariusz.Gora@desy.de}}
        
\author{Elisa Bernardini\\
       Institut f\"ur Physik, Humboldt Universit\"at, Newtonstr. 15,  D-12489 Berlin, Germany\\
       Deutsches Elektronen-Synchrotron (DESY), Platanenallee 6, D-15735 Zeuthen, Germany\\
        E-mail: \email{Elisa.Bernardini@desy.de}}

\abstract{This paper investigates the potential to detect tau neutrinos in the energy range of 1-1000 PeV searching for very inclined showers with imaging Cherenkov telescopes. A neutrino induced tau lepton escaping from the Earth may decay and initiate an air shower which can be detected by a fluorescence or Cherenkov telescope.  We present here a study of the detection potential of Earth-skimming neutrinos taking into account neutrino interactions in the Earth crust,
local matter distributions at  various detector sites, the development of tau-induced showers in air and the detection of Cherenkov photons with IACTs. We analysed simulated shower images on the camera  focal plane and implemented generic reconstruction chains based on Hillas parameters. We find that  present IACTs can  distinguish  air showers induced by  tau neutrinos  from  the  background of hadronic showers in the PeV-EeV energy range. We present  the neutrino 
trigger efficiency obtained for a few configurations being considered for the next-generation Cherenkov telescopes,  i.e. the Cherenkov Telescope Array.  Finally, for a few representative neutrino  spectra expected  from astrophysical sources, we compare the expected event  rates at running  IACTs to what expected for the dedicated IceCube neutrino telescope.}

\FullConference{The 34th International Cosmic Ray Conference,\\
		30 July- 6 August, 2015\\
		The Hague, The Netherlands}

\begin{document}
\section{Introduction}
The  existing Imaging Air Cherenkov Telescopes (IACTs) such as MAGIC~\cite{magic}, VERITAS~\cite{veritas} and H.E.S.S.~\cite{hess} could have the capability to detect PeV tau neutrinos by searching for very inclined showers~\cite{fargion}. In order to do that, the Cherenkov telescopes need to be pointed in the direction of the taus escaping from the Earth crust, i.e.\ at or a few degrees below the horizon. In \cite{upgoing_magic}, the effective area for up-going tau neutrino observations with the MAGIC telescopes was calculated analytically and found to be maximum  in the range from 100\,TeV to $\sim$~1\,EeV. However, the  sensitivity for diffuse neutrinos was found to be  very low because of the limited FOV (the topographic conditions allow only for a small window of about 1 degree width in zenith and azimuth to point the telescope downhill),
the short observation time and the low expected neutrino flux.

On the other hand, if flaring or disrupting point sources such as GRBs are observed, one can  expect an observable number of events even from a single GRB if close by, as recently  shown  by    the All-sky Survey High Resolution Air-shower (Ashra) team \cite{Asaoka:2012em}. Also, for  IACT sites with different topographic conditions, the acceptance for up-going tau neutrinos is increased by the presence of mountains~\cite{gora:2015}, which serve as target for neutrino interaction
 leading to an enhancement in the flux of emerging tau leptons. A target mountain can also shield against cosmic rays and star light. Nights with high clouds offen prevent the observation of $\gamma$-ray sources, but still allow  point the telescopes to the horizon. While observation of tau neutrinos is not the primary goal of IACTs, a certain level of  complementarity can be expected where  switching for normal (i.e. $\gamma$-ray) observation made to tau neutrinos (i.e. mostly horizontal) pointing.
 Next-generation Cherenkov telescopes, i.e.\ the Cherenkov Telescope Array (CTA)~\cite{cta}, can in addition exploit their much larger FOV (in extended observation mode) and  a higher effective area.

\section{Method}
\label{method}

In order to study the signatures expected from neutrino-induced showers by IACTs, a
full Monte Carlo (MC) simulation chain was set, which consists of three steps.
First, neutrino propagation of a given neutrino flux through the Earth and the atmosphere is simulated using  an extended version of the ANIS code~\cite{gora:2007}, see also~\cite{gora:2015} for more details. Then,   the shower development  of $\tau$-induced showers  and  Cherenkov light production  from such shower  is simulated with CORSIKA~\cite{corsika}.  The CORSIKA   (version 6.99)   was compiled with TAULEP option~\cite{taulep}, such that the tau decay is simulated with  PYTHIA package~\cite{pythia}. In order  to simulate  Cherenkov light from  inclined showers for any defined Cherenkov  telescopes array the  CERENKOV and  IACT option was also activated~\cite{simtelarray}. Finally,  to consider  the atmospheric depth correctly for inclined showers, the  CURVED EARTH and SLANT option was also selected.

For high energies ($>1$ PeV)  the computing time become excessively long (scaling roughly with the primary energy). In order to reduce it to tolerable values the so-called
"thin sampling" mechanism is used~\cite{thinning}. To cope with the vast number of secondary particles  thinning and re-weighting of secondaries was used with a thinning level of 10$^{-6}$ . The kinetic energy thresholds for explicit tracked particles were set to: 300, 100,  1,  1~MeV for hadrons, muons electrons and photons, respectively. Shower simulations were performed considering  QGSJET II  model  for hadronic interactions in the atmosphere. 

We simulated showers induced by  tau leptons with energies from  1 - 1000 {PeV} in steps of 0.33 decades and   with an injection positon at altitudes ranging from   detector  level
 \footnote{ We used 1800 a.s.l for the simulation of current generation  of  IACTs  and  2000 m a.s.l  for  CTAs.} to the  top of atmosphere. The injection point spans  different vertical depths  from  ground to top of the atmosphere with steps of at least  50 g/cm$^2$.  At vertical depth,  1000 showers  were generated in order to  study shower-to-shower fluctuations and to cover  different tau decay channels. 
\begin{figure*}[t]   
\vspace{-0.5cm}                                                                                                          
\begin{center}                                                                                     
\includegraphics[width=0.45 \columnwidth]{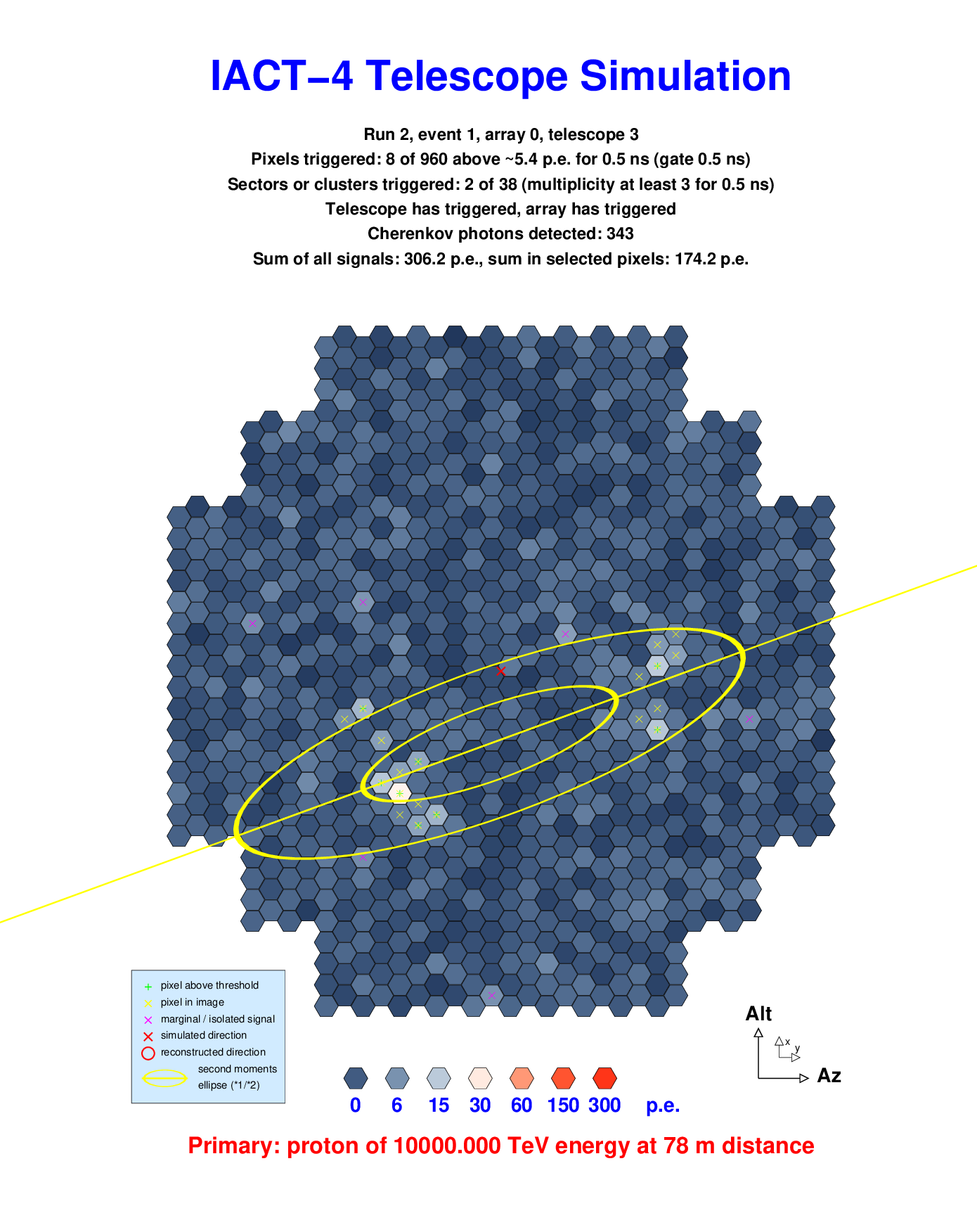}    
 \includegraphics[width=0.45 \columnwidth]{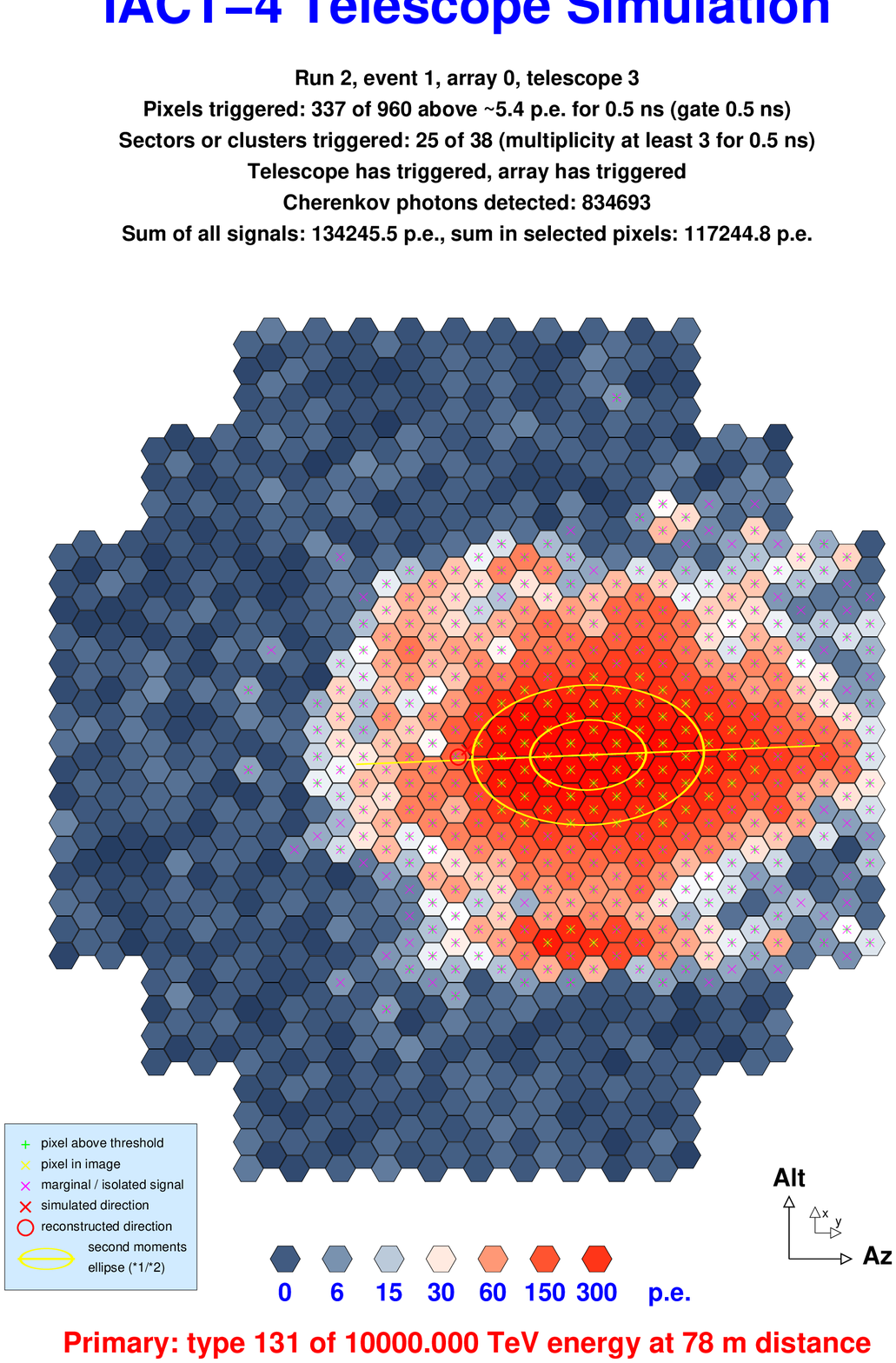}
\end{center}
\vspace{-0.5cm}
 \caption{\label{fig::hessimage} Example of simulated shower images with primary partilcle energy 10 PeV  and zenith angle $\theta=88^{\circ}$ as seen by IACT-4 camera with 2.5$^{\circ}$ FOV. (Left)  
 proton interacting at the top of atmosphere,   first interaction point at vertical depth below 50 g/cm$^2$ and detector-to-shower distance of about  1000 km; (Right) tau  lepton  tau decaying close to the detector, with an  injection vertical depth of  760 g/cm$^2$ and a  detector-to-shower  distance  of about  50 km.} 
\end{figure*}   
The results of CORSIKA simulations were used as  input for the last step i.e. simulation of  detector response. We used  the Cherenkov telescope simulation package: {\tt sim\_telarray} ~\cite{simtelarray} . The light collection area is simulating  including the ray-tracing of the optical system, the measured transmittance and the quantum efficiency of PMT. The response of the  camera  electronics  was simulated in detail  including   night-sky background and different system triggers.
The {\tt sim\_telarray} simulations were performed for different configurations: H.E.S.S. like   four telescopes 
(named here by  IACT-4), and  for  a few  CTAs arrays  considered  in~\cite{ctasim} with so-called {\it production-1} settings. 
The IACT-4   can be considered as representative for current generation of IACTs. The  response to   $\tau$-induced showers
is found to depend weakly on the details of the optical set-up, field of view and  camera electronics.   Among  different CTAs array configurations  shown in~\cite{ctasim}  the arrays  chosen were  named CTA-E  (59 telescopes)
and CTE-I  (72 telescopes),  which according to ~\cite{ctasim} are the best compromise between compact   and  dense layout. The selected arrays  have only slightly worse  sensitivity for $\gamma$-rays  than the  full  CTA array~\cite{ctasim}.  

In order to compare images at the camera plane we also simulated inclined showers induced by  proton, photon  and  electron. At  energies larger than  1 PeV, we do not expected significant background of showers initiated by photons or electrons. The proton  simulations were instead  used to estimate the main isotropic background for  neutrino searches due to   interaction of comics rays  in atmosphere.  The direction of primary protons was varied   within a circular  with aperture   $\beta=5^{\circ}$ around the fixed primary direction, i.e. the VIEWCONE option was selected in the CORSIKA simulations.

\section{Results} \label{sec:results}

In case of   showers observed at large zenith angles the Cherenkov telescopes 
 has to undergo a long optical path, due to  a thicker layer of atmosphere. The shower maximum is located far
 from the observatory and the photon density  at the mirrors decreases. This reduces the efficiency compared to lower zenith angles, especially at low energies. Images on the camera  will be dimmer and smaller in size. As an example, in Figure~\ref{fig::hessimage}   we show  a representative  shower image for a 10~PeV proton  injected at the top of atmosphere and 10 
 PeV tau lepton injected  close to the detector, respectively.  As expected, the shower image on focal camera plane for  tau  has much large image  size  and contains much more photons  comparing to the  proton one. Note also, that for inclined showers
the hadronic and electro-magnetic component is  almost completely  absorbed in the atmosphere while the muonic component (muons) can reach the Earth. Thus, the showers images  on the cameras from proton-induced showers  will be mostly contain the  muon ring (if muons propagating parallel to the optical axis) or   incomplete ring (arcs) in the camera, see Figure~\ref{fig::hessimage} (Left) as an example.

\begin{figure}[t]   
   \vspace{-0.5cm}                                                                                                       
  \begin{center}                                                                                     
    \includegraphics[width=0.49\columnwidth,height=6.0cm]{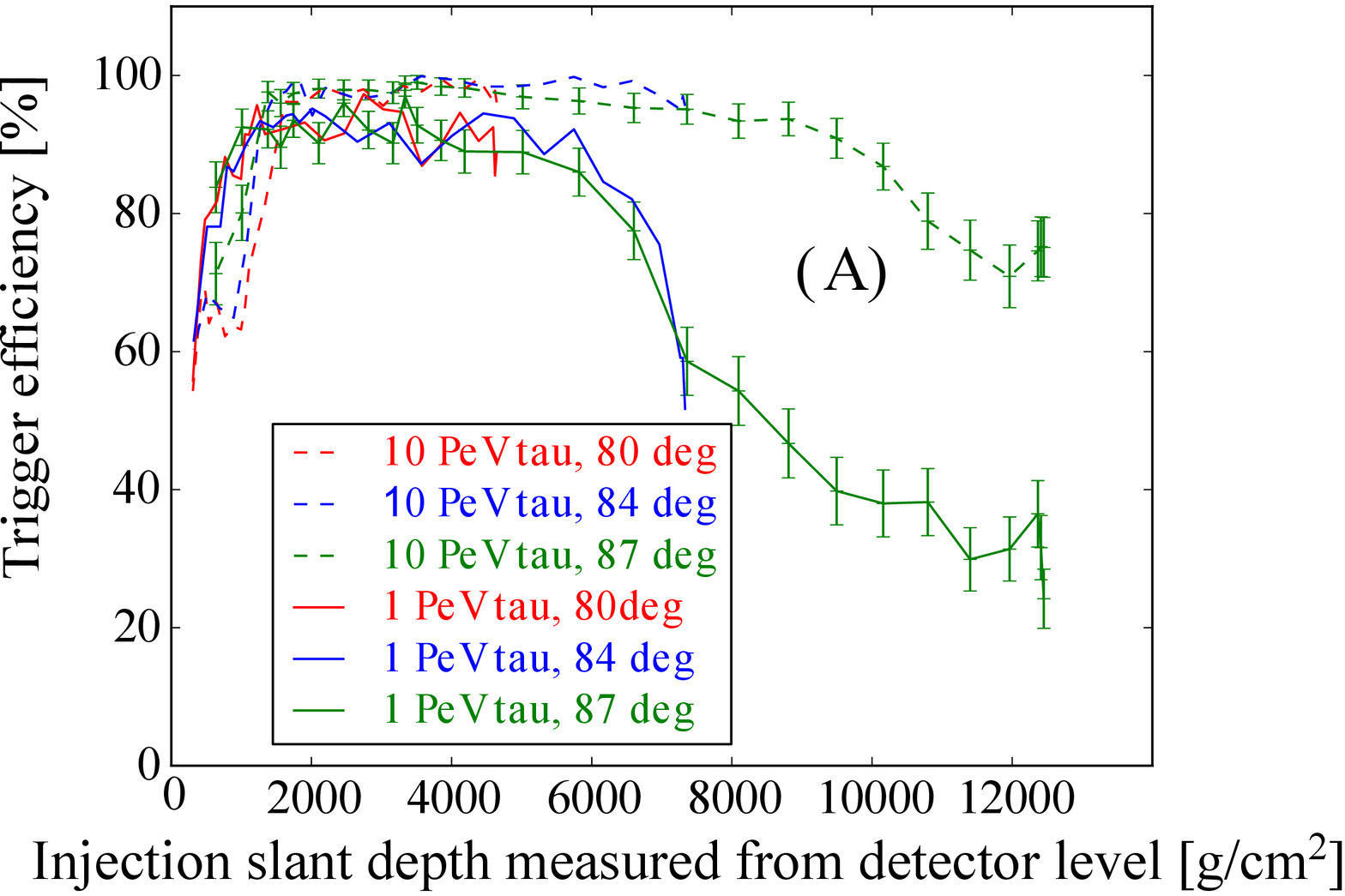}  
      \includegraphics[width=0.49\columnwidth,height=6.0cm]{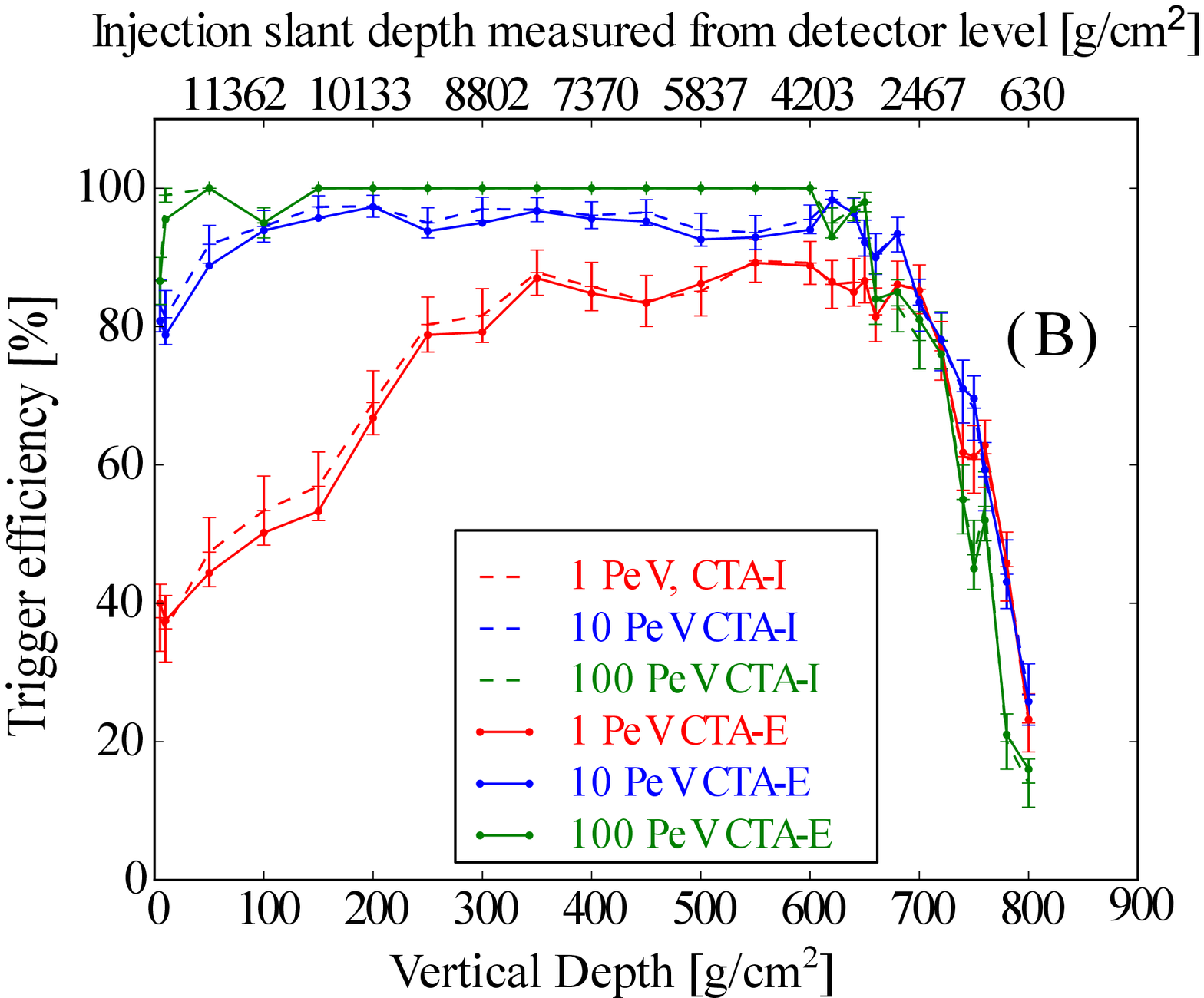}  
  \end{center}
\vspace{-0.5cm}
 \caption{\label{trigger333}(A) Trigger efficiency as a function of  injection slant depth  with IACT-4  for different zenith angles and energies of tau lepton.  Note, that for different zenith angles   the distance from atmospheric border to detector level  is significant different due to the Earth's curvature.  (B) Trigger probability for  CTAs at a fixed zenith angle of  $87^{\circ}$. }  
 \vspace{-0.5cm}
\end{figure}

Figure~\ref{trigger333} (A) shows  the trigger  probability~\footnote{ It is defined 
as the  number of simulated showers with positive trigger decision over the total number of generated showers.
In this work,   simulation was  done  for two level  trigger, so-called Majority trigger. The first level  is a camera level  trigger ({\bf L1}) defined by  3 pixels  above 4 photo-electrons (p.e.) with a short time window and the second level is basically a coincidence level trigger amoung all telescopes in the  defined array or sub-array ({\bf L2}) and requires at least 2 neighboring  triggered telescopes
}  for $\tau$-induced showers with different zenith angles and  energies of tau lepton in case of IACT-4 array. The  calculated trigger probabilities for different zenith angles $\theta=80^{\circ}, 84^{\circ}, 87^{\circ}$  are  quite similar, within errors,  if their  plotted as a function of the  distance between the injection point and the detector  measured in g/cm$^2$. This is understood, if we note  that  amount of  Cherenkov light detected by telescopes   depends essentially on atmospheric slant depth interval  between the Cherenkov telescope and  the shower maximum.  At shower maximum  shower has the largest lateral extension and  Cherenkov light productions,  thus is capable of producing the largest signal  seen by IACTs  telescopes. 
As we now,  could not simulate the showers with  zenith angle $\theta>90^{\circ}$ in the combination of "CURVED EARTH" and IACT options, we use the zenith angle $87^{\circ}$ to estimate the trigger efficiency for up-going tau neutrino showers. This should be reasonable assumption, because  the trigger efficiency  in case of  $\tau$-induced showers with the same energy  should only slightly depends on zenith angle (as its confirmed by Figure~\ref{trigger333} (A)),  as long as the corresponding altitudes of  shower maxima  are the similar.

 As expected (see Figure~\ref{trigger333} (A)) the trigger probability  increases with  primary energy of lepton tau and decreasing distance to the detector. Only,   at particle  injection slant depths $<1000$ g/cm$^2$, measured from detector level,  the trigger efficiency drops due to fact that shower maximum is to close to the detector or the shower did not reach yet the maximum of shower development,  decreasing amount of Cherenkov light seen by telescopes. It is also worth to mention, that
below  6000 g/cm$^{2}$     the trigger probability  is  at the level of about  90\%. In this case the  corresponding geometrical  distance to the detector (in meters) depends on  zenith angle $\theta$, but  for $\theta=87^{\circ}$  is of  about   $\sim100$ km. This  gives estimate  about the size of the  active volume for $\tau$-induced showers seen by IACTs.  
\begin{figure*}[ht]   
\vspace{-0.5cm}                                                                                                             
  \begin{center}                                                                                      
     \includegraphics[width=\columnwidth]{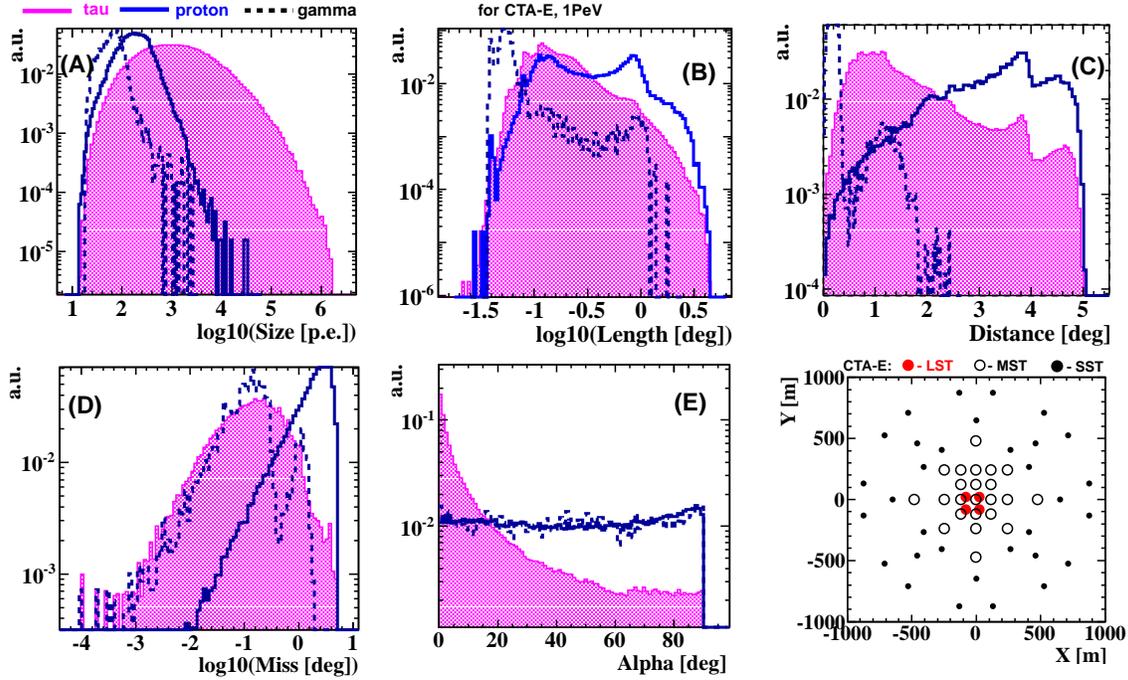} 
  \end{center}
  \vspace{-0.5cm} 
 \caption{\label{fig::hillas} (A-E) Normalized distribution of Hillas parameters  for $\tau$, $p$ and $\gamma$-induced showers  and primary particle energy 1PeV,  zenith angle $\theta=87^{\circ}$ and CTA-E. Only  deep $\tau$-induced showers with  
 injection depth  larger than 300 g/cm$^2$ are shown, while for $p/\gamma$ only events  interacting at the top of atmosphere with the first interaction point < 100 g/cm$^{2}$ (vertical depth)  are considered; (F) The   CTA-E  layout   
 considered  in this work. The array consist of  59  telescopes,   of different  size i.e.  Large Size Telescopes (LST) with
  $\sim23$ m aperture, $5^{\circ}$ FOV and  $0.09^{\circ}$ camera pixel size  (red full circles),  Medium Size Telescopes (MST) 
  with $\sim$ 12 m aperture,  $8^{\circ}$  FOV and $0.18^{\circ}$ camera pixel size  (open black circles) and
   Small Size Telescopes  (SST) with $\sim$ 4-7 m aperture,  $10^{\circ}$  FOV and  $0.25^{\circ}$ camera pixel size (black full circles).   For more detail description of telescope properties, see Table 1 in~\cite{ctasim}. } 
 \vspace{-0.3cm}  
\end{figure*}   
 
  Figure~\ref{trigger333} (B) shows  the trigger probability for  considered   CTAs arrays and  different  primary  energy of lepton tau.  As  for IACT-4 array,  the trigger  probability  is close  increases because the higher the energy, the more Cherenkov light is produced, and the larger the number of triggered events. It is  also well seen by comparing with results from  Figure~\ref{trigger333}~(A), that  calculation for larger CTAs array,  which consists of  much more telescopes with different optics and camera structures,
 than  IACT-4 array gives  basically similar  fraction of triggered events. Moreover, for  considered  CTAs arrays,  the trigger efficiency  only  slightly dependence on  array  structure.  This  can be explain by the fact, that for inclined shower  studied in this work (with $\theta>80^{\circ}$)   the  size  of Cherenkov light pool distribution  at detector level  is  larger than 1~km~\footnote{For index of refraction $n_{air}=1.00023$ at an altitude of 1800~m, the Cherenkov opening angle is $\alpha \simeq 1.2^{\circ}$.  Thus, for geometrical distance from the  shower maximum to detector
of  about   50 km the Cherenkov ring  radius on the ground,  assuming  not changes of refraction index within this distance, is given by:  50~km~$\times \tan(\alpha)/ \cos(\theta)$=1.04 km/{$\cos(\theta$)} km  for fixed zenith angle $\theta$.}, which  is much more  than  distance between telescopes in considered arrays. Thus, the fraction of triggered events is expected to be similar and only slightly depends on density of telescopes.
 
The Cherenkov light forms an ellipse on the cameras. 
The cleaned camera image is characterized by a set of image parameters  based on Hillas~\cite{hillas}.
 These parameters provide a geometrical description of the images of the showers and are used to infer the energy of the primary particle, its arrival direction and to distinguish between gamma-ray showers and hadronic showers. It is interesting to  study  distributions of these parameters also  in the case of  deep $\tau$-induced showers, see Figure~\ref{fig::hillas} (A-E).  As we can see from plots,  the 
distribution of Hillas parameters for deep $\tau$-induced showers  are quite different than corresponding  distribution for $p$ and $\gamma$-induced shower developing at the top of atmosphere. In general, these parameters  depend on the  geometrical distance of shower maximum to the detector, which for deep $\tau$-induced shower is much smaller than for inclined  $p$ and $\gamma$-induced showers  developing  at  the top of atmosphere. For example, at  $\theta>80^{\circ}$  this distance is of about a few  hundred kilometers for particle interacting at the top of atmosphere and only a few tens kilometers for deep  $\tau$-induced shower.  Thus, this geometrical effect leads to  rather  good separation of close ($\tau$-induced) and far-away ($p$, $\gamma$) events in the Hillas parameter phase space.

We  found that the shape  of   $Distance$, $Miss$ and $Alpha$ distributions only slightly depend on primary particle energy   and  shower zenith angle  (above  $80^{\circ}$). Hovewer, as expected, for  energy dependent parameters like: $Size, Lenght, Width$   the  shift  of  maximum of corresponding distributions to the higher values was observed,  when simulation was done for the higher primary particle energies. It is also worth to mention, that  the largest  differences in Hillas distributions between deep $\tau$-induced showers and  $p$ and $\gamma$-induced  showers are present for $Size$, $Miss$ and $Alpha$ parameter.  This  gives possibility to use these observables  in order  to  distinguish deep $\tau$-induced shower from the background of inclined hadronic showers. 
\begin{figure}[t]   
\vspace{-0.5cm}                                                                                                          
  \begin{center}                                                                                  
    \includegraphics[width=0.49\columnwidth,height=5.5cm ]{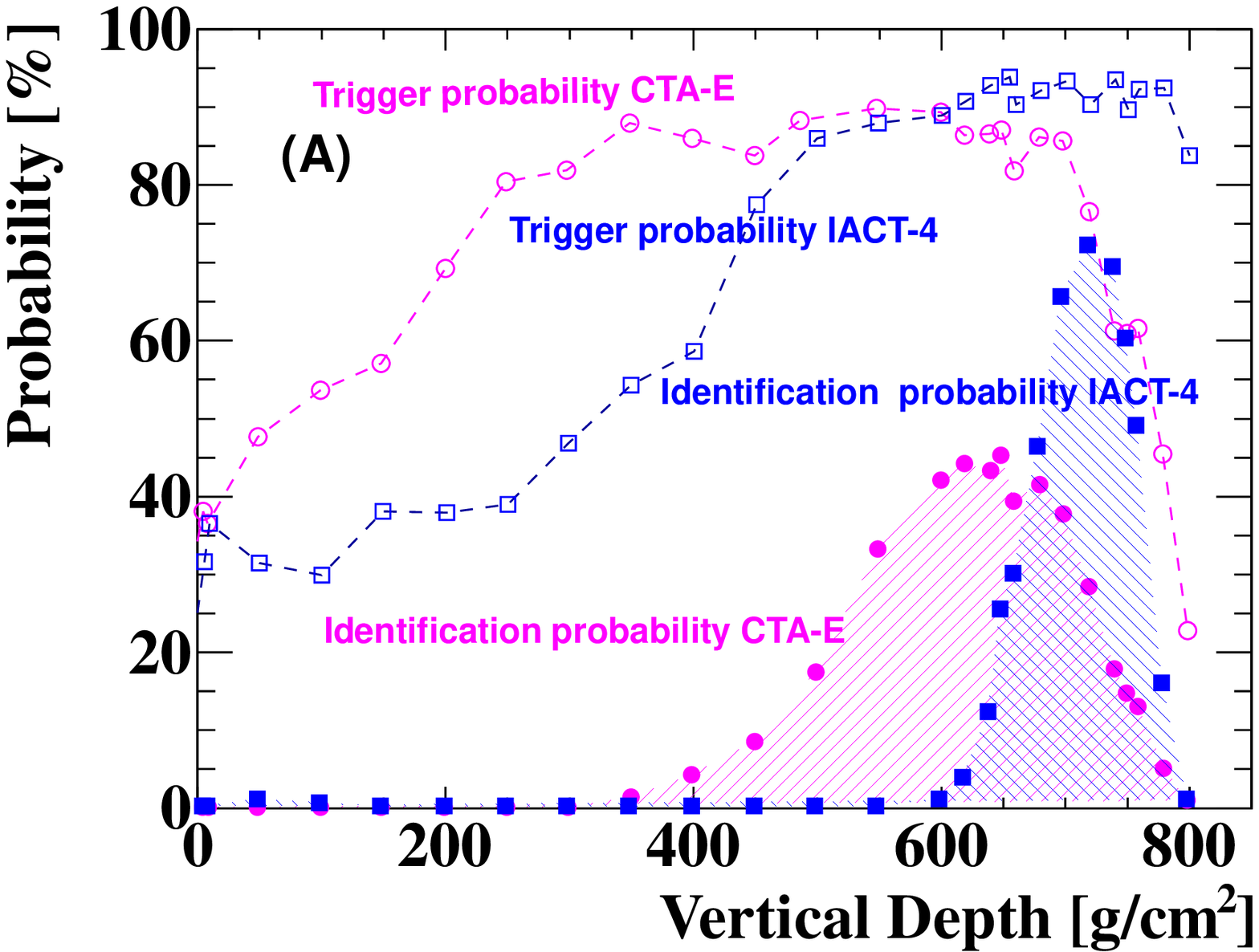}   
     \includegraphics[width=0.49\columnwidth,height=5.5cm]{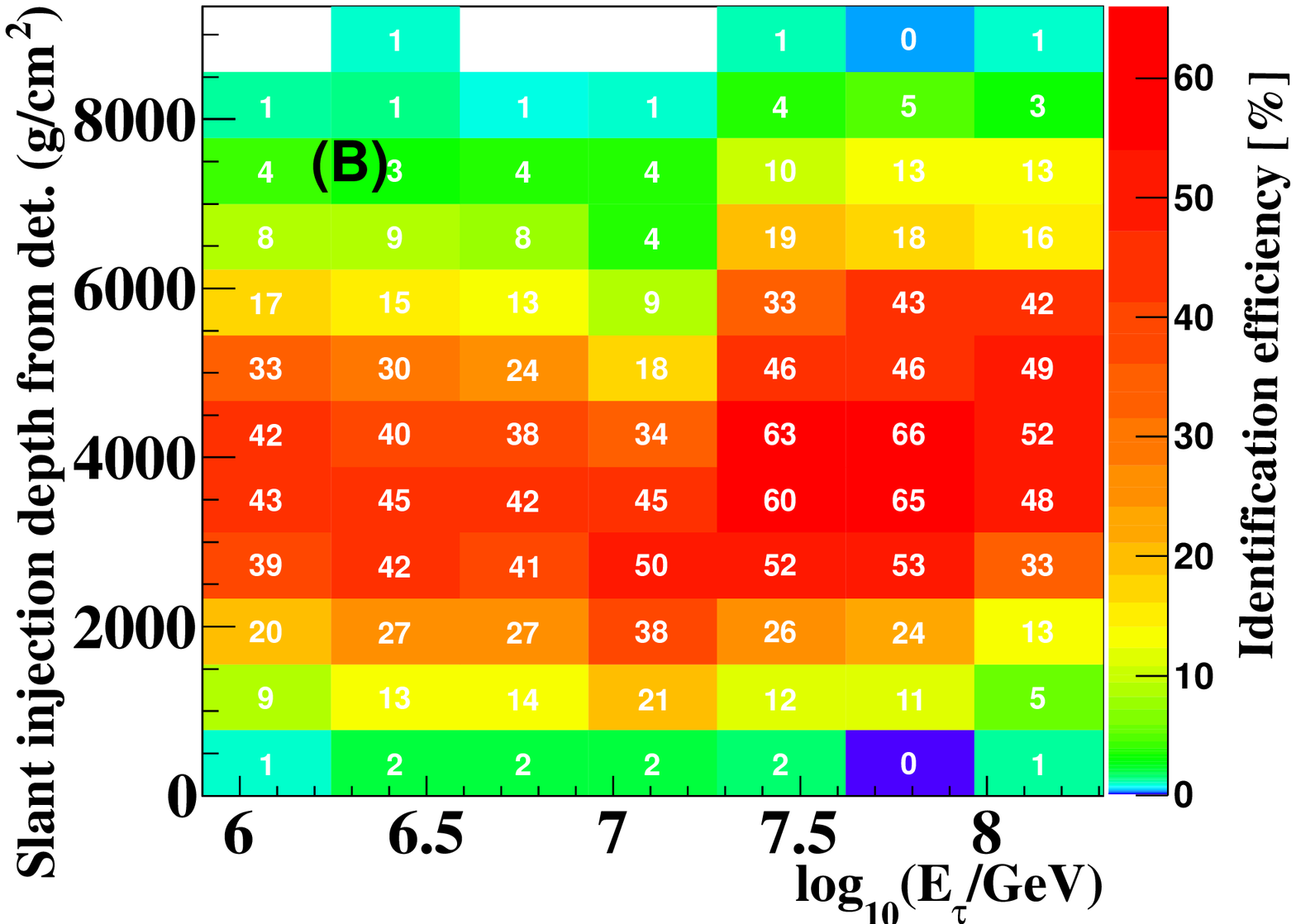}
  \end{center}
\vspace{-0.4cm}
 \caption{\label{trigger3333} (A) Trigger and identification efficiency for 1 PeV $\tau$-induced shower  with zenith angle $\theta=87^{\circ}$ for IACT-4 and CTA-E, (B) Identification efficiency for CTA-E as a function of lepton tau energy and injection slant depth measured from detector level.}  
 \vspace{-0.4cm}
\end{figure}  

In order to evaluate the best set of cuts to identify deep $\tau$-induced neutrino showers, we used the program GARCON~\cite{garcon}, which is based on genetic algorithms  to optimize cuts in a way that the signal passing rate is maximized while the background contamination is minimal. In this work, the fifth parameter phase space 
$\vec{x} = \{Size, Length, Distance, Miss, Alpha\}$ was used to maximize the functional $F[S(\vec{x}^{cut}),B(\vec{x}^{cut})] = S(\vec{x}^{cut})/\sqrt{ B(\vec{x}^{cut})}$,  where $S(\vec{x}^{cut})$ is the number of  deep $\tau$-induced showers ($\tau$ injection depth larger than 600 g/cm$^2$ i.e. $\sim 43$ km from detector for $\theta=87^{\circ}$) passing after cuts and $B(\vec{x}^{cut})$ the number of remaining $p$-induced shower
at top of atmosphere ($p$ injection depth < 100 g/cm$^{-2}$) after cuts.
The  set of optimized cuts on the identification observables,  which lead to 0 proton events and signal efficiency of about 30\%, are  listed  in  Table 1 for 
IACT-4 and CTA-E.

In Figure~\ref{trigger3333} (A) the  influence of cuts on the trigger probability are shown, while in Figure~\ref{trigger3333} (B) 
 we present the dependence of  identification efficiency  as a function  of  primary energy  of lepton tau.  It should be pointed out, that due to larger number of triggered proton events   (the larger background level for CTA-E), the optimized cuts  in this case 
 are usually  harder (see Table~1) than for IATC-4. This   leads to the  smaller  values of identification efficiency for  CTA-E  than IACT-4 array,
  at  vertical depth larger than 600 g/cm$^2$. However,   as expected the  distribution is extended to the lower values of injection depth, up to 400 g/cm$^2$. In addition, the difference in the trigger/identification efficiency  seen for  vertical depth > 750 g/cm$^{-2}$ between IACT-4 and CTA-E it  is cost by  the different detector  altitudes. The difference  is only  200 m, but  for zenith angle $\theta=87^{\circ}$  it    translates to  of about 4 km difference in detector to shower distance, which in case of IACT-4  leads to the larger fraction of showers which already reached the maximum of shower development. 
 \begin{table}[hb]
\vspace{-0.0cm}
\small
\center
\begin{tabular}{ccccccccc}
\hline 
array&E$_{i}^{\tau}$ & $Size$ &  $Length$& $Distance$ & $Miss$ & $Alpha$ &  $S(\vec{x}^{cut})$ & $ B(\vec{x}^{cut})$ \\ 

type&[PeV]  & [p.e.] & [deg] & [deg] & [deg] & [deg] & [\%] & [\%] \\ 
\hline 
\hline
IACT-4& 1& > 3110  & < 0.46  & < 1.09  & < 0.18   & < 45.1  & 33   & 0 \\ 
 CTA-E && > 791  & <  0.27 & < 2.91 & <  2.12  & < 13.8 & 24 & 0 \\     
   & &   &   &   &   &   &  &  \\  
IACT-4&10 & > 3.55E3 & < 0.70  & < 2.19 & < 0.30 & < 48.7  & 31 & 0 \\ 
CTA-E & & > 3.51E3& <  0.26 & < 2.97  & < 0.56 & < 41.4 & 34 & 0\\ 
& &   &   &   &   &   &  &  \\ 

%
IACT-4 &100 & > 4.31E4  & < 0.78 & < 1.89 & < 1.81 &  < 12.7& 30  & 0\\ 
CTA-E & & > 1.81E4  & < 0.33 & < 2.93 & < 0.22 &  < 13.5& 33  & 0\\ 
\hline 
\end{tabular} 
\caption{Final cuts for  identification observables obtained from GARCON optimization for 
$\tau$-induced showers with $\theta = 87^{\circ}$. $E_{i}^{\tau}$ stands
for initial injected particle energy.} 
\vspace{-0.5cm}
\end{table}
  As it also seen from Figure~\ref{trigger3333} (A)  for IACT-4 an  average   identification efficiency is  at level of  ~30\%  for lepton tau  with energy of  1 PeV.  Comparing this number to the value used in our previous work~\cite{gora:2015},  where an average trigger efficiency of $\sim 10\%$ was assumed, we see that more realistic trigger simulations gives
at least  3 times larger value.  As a consequence, this also  leads to increase of  event rate  calculated  for $\nu_{\tau}$  in our recent work, about the  similar factor, see Table 1 in \cite{gora:2015} for more details.
\section{Conclusions} \label{sec:conclusion}

 In this paper,  we present results of MC simulations  of   $\tau$-induced air showers  for IACTs  and  for selected  CTA  arrays.  We  calculated the  trigger and identification efficiencies  for $\tau$-induced showers and  study  properties of  images for $\tau$-induced shower  on camera  focal plane,  as described by Hillas parameters.   In  our previous work~\cite{gora:2015}, we show, that  the  calculated  neutrino  rates  are  comparable or even larger (above $\sim30$ PeV)  to  what  has  been  estimated  for  the  IceCube neutrino telescope assuming realistic observation times for  Cherenkov  telescopes  of  a  few  hours.  However previous calculation was done for ideal detector. Instead here, we show that  more realistic simulation will be lead  to the  larger  even rate as seen by IACTs/CTAs, which make observation of $\tau$-induced shower by  the present  or future Cherenkov telescopes more promising.

\end{document}